\documentstyle[amssymb,12pt]{article}
\begin{document}
\renewcommand{\theequation}{\thesection.\arabic{equation}}
\begin{titlepage}

\begin{center}
{\Large\bf  Decay of the Two-Point Function in One-Dimensional
$O(N)$ Spin Models with Long Range Interactions}
\bigskip\bigskip\bigskip\\
Herbert Spohn\bigskip\\
Zentrum Mathematik, Technische Universit\"at M\"unchen,\\
D - 80290 M\"unchen, Germany\\
email: {spohn@mathematik.tu-muenchen.de}\\

\bigskip\bigskip\bigskip

Wilhelm Zwerger\bigskip\\
Theoretische Physik, Ludwig-Maximilians-Universit\"at,\\
Theresienstr. 37, D - 80333 M\"unchen, Germany\\
email: {zwerger@stat.physik.uni-muenchen.de}\\
\end{center}

\vspace{5cm}
\noindent
{\bf Abstract}. Using Griffiths and Lieb-Simon type inequalities, it is shown
that the two-point function of ferromagnetic spin models with $N$ components
in one dimension decays like the interaction $J(n) \sim n^{-\gamma}$
provided that $1 \le N \le 4$ and $T > T_c$.
\end{titlepage}

\section{Introduction and Main Result}
\setcounter{equation}{0}

As is well known, classical spin models with a continuous symmetry
in two dimensions lead to scale invariant field theories with
the nonlinear sigma-model action 
\begin{equation}
{1 \over {2T}} \int d^2x (\nabla {\bf
s}({\bf x}))^2,  
\end{equation}
where ${\bf s}$ is a unit spin with $N$ components. The resulting behaviour 
distinguishes between an abelian, $N=2$ plane rotor, and a
non-abelian symmetry group, $N \ge 3$. In the latter case the two-point correlation,
$g({\bf x}) = <{\bf s} ({\bf x}) \cdot {\bf s}(0)>$ decays exponentially at any finite
temperature with a finite correlation length $\xi(T) \sim \exp[2 \pi/(N-2)T]$
\cite{P,BZ}. On the other hand for an XY-symmetry, the exponential decay holds only
above the Kosterlitz-Thouless critical
temperature $T_{KT}$. The phase at low
temperatures exhibits power-law decay $g({\bf x}) \sim |{\bf x}|^{- \eta(T)}$
with a continuously varying exponent $\eta(T)$ \cite{KT}.  Qualitatively this
behaviour can be understood through the spin-wave approximation
\begin{equation}
(\nabla {\bf s})^2 \approx (\nabla {\bf \varphi})^2 
\end{equation}
with ${\bf s} = (\cos \varphi,\sin \varphi)$,
neglecting the periodicity of the phase variable $\varphi$. Since the approximate
action is Gaussian, the correlation function
$g({\bf x}) = \Re <\exp [i(\varphi({\bf x}) - \varphi(0))]>$ can be calculated
easily, yielding the power law decay $|{\bf x}|^{- \eta(T)}$ with
$\eta = T /2 \pi$. The vortex
excitations lead, at low $T$, only to a finite
renormalisation of $\eta$ \cite{KT}.

In one-dimensional models the spatial dimension can be mimicked
by a long range interaction with a decay as $|n|^{- \gamma}$, $n$ 
being a point on the one-dimensional lattice.
One notices that for $\gamma = 2$ the action is again
scale invariant. For this marginal case, it was conjectured early by Thouless
\cite{T} that the $N=1$ Ising model has a spontaneous magnetization $m^*$ below a
nonzero critical temperature $T_c$. The spontaneous magnetization
jumps to a finite value at $T_c$, yet the 
transition is
continuous. This conclusion  was confirmed through an analysis of the 
equivalent Kondo problem
\cite{AYH,AY} and later proven rigorously \cite{FS,ACCN}. A renormalization group
calculation for long range spin models in one dimension both for $N=1$ and
a continuous symmetry was performed by Kosterlitz \cite{K}. Within a one loop
calculation he showed that there is always a low temperature 
spontaneous magnetization provided $1<
\gamma <2$ $(\gamma > 1$ being necessary to have an extensive free energy). In the
marginal case $\gamma = 2$ and for $N \geq 2$, the associated beta function
vanishes quadratically near the trivial fixed point $T =0$. This indicates
that $T_c = 0$ for $N \geq2$ and an exponential behaviour $\chi(T) \sim
\exp[2 \pi^2/(N-1)T]$ for $T \rightarrow 0$ of the 
susceptibility. Due to the power law interaction there can be no
finite correlation length, however.

In our present note we discuss the long distance behaviour  
of the two-point function $g(n)$. In the phase where $m^* =0$ it is shown
that if $g(n)$ has a power law decay at all, it is necessarily equal to that
of the interaction. In particular, a spin wave approximation
is qualitatively incorrect for long range models even at very low
temperatures. Moreover, in the XY-case with marginal $\gamma = 2$, our
rigorous bounds rule out the appearance of a low temperature phase with a 
continuously varying exponent
$\eta(T)$ and infinite susceptibility, which
has been claimed in the literature on the basis of spin wave theory and Monte
Carlo simulations \cite{S,BS,SF}. Our result is of direct relevance 
to the problem of strong tunneling in the so called 
single electron box, showing that the Coulomb blockade at zero 
temperature is not destroyed even for large conductance \cite{HZ}.

To be more precise, we consider the spin Hamiltonian
\begin{equation}
H = - {1 \over 2} \sum_{m,n} J(m-n) {\bf s}_m \cdot {\bf s}_n,
\end{equation}
where the couplings are ferromagnetic, $J(n) \ge 0$, and decay as
\begin{equation}
J(n) \cong |n|^{- \gamma}
\end{equation}
for $|n| \rightarrow \infty$. As before the two-point function is
defined by $g(n) = <{\bf s}_n \cdot {\bf s}_0>$ in the infinite
volume limit with free boundary conditions. Then
\begin{equation}
\lim_{n \rightarrow \infty} g(n) = m^{*2},
\end{equation}
where $m^*$ is the spontaneous magnetization with the standard
convention that $m^* = 0$ for $T >
T_c$ and $m^* > 0$ for $T < T_c$. We define the scaling exponent $\eta$
by
\begin{equation}
g(n) - m^{*2} \cong |n|^{-\eta}
\end{equation}
for large $n$. The magnetic susceptibility is given by
\begin{equation}
\chi = {\beta \over N} \sum_n(g(n) - m^{*2}).
\end{equation}
If $\eta < 1, \ {\rm then} \ \chi = \infty$.

The qualitative phase diagram for such feromagnetic models is rather 
well understood. For 
$\gamma > 2$ one has $m^{*} = 0$  and for 
$\gamma < 2$ $m^{*} > 0$ at sufficiently low temperatures. In the marginal case, $\gamma = 2$, the number 
of components becomes relevant. Whereas for the Ising model, $N=1$, 
$m^{*} > 0$ at low $T$  \cite{FS}, for $N\geq 
2$ Simon \cite{S1} proves that $m^* = 0$
 at any finite $T$. The decay of the two-point function 
has been studied on a rigorous level mostly for the Ising model with 
particular 
attention to the marginal case $\gamma = 2$ \cite{I}: For $T > T_c$ one has $\eta = 2$. At $T_c \ m^*$
jumps to a non-zero value, the Thouless effect, and $\eta$ jumps to
zero. Below $T_c$, $\eta(T)$ increases with decreasing $T$ and locks to its high 
temperature value $\eta =2$
at some critical value $T_c^* < T_c$. Although $\eta$ varies
continuously, the overall behavior is obviously quite distinct from the
standard Kosterlitz-Thouless scenario in the short range $d=2, N=2$ case.

Here we show that if $N=1,2,3,4$, if
$m^*=0$, and if $g(n)$ is known to have some decay already, then
$\eta = \gamma$. A lower bound of this form is
known from Griffiths second inequality for arbitrary $N$ \cite{G}. A
corresponding upper bound is slightly more involved. It uses a
Lieb-Simon type inequality \cite{S2,L}, which relies on Gaussian domination
of the four-point function. Although this is expected to hold in
general, it has been proved only for $N=1,2,3,4$ components \cite{B}.

In the following section we give the details of the argument. In
fact, it would be of interest to have a numerical solution of the
nonlinear integral equation (2.8), which could be used as a sharp
test of Monte-Carlo simulations.

\section{Bounds on the two-point function}
\setcounter{equation}{0}

We consider ferromagnetic spin models with $N$ components in one
space dimension with Hamiltonian (1.3).
%\begin{equation}
%H = - {1 \over 2} \sum_{m,n} J(m-n) {\bf s}_m \cdot {\bf s}_n.
%\end{equation}
Here ${\bf s}_n$ is the $N$ component spin at lattice site $n$,
$n$ \ integer, with $|{\bf s}_n|=1$. The couplings $J$ satisfy $J(n) \ge
0, J(n) = J(-n)$ and have the asymptotic decay (1.4).
%\begin{equation}
%J(n) \cong c |n|^{-\gamma}
%\end{equation}
%for large $n$. 
To have an extensive free energy we require $\gamma >
1$. The equilibrium distribution in finite volume $[-\ell,...,\ell]$ is
given by
\begin{equation}
Z^{-1} \exp[-\beta H] \prod^{\ell}_{n=-\ell} \delta(|{\bf s}_n|-1)d^Ns_n.
\end{equation}
We choose free boundary conditions, i.e. ${\bf s}_n=0$ for $n$
outside $[-\ell,...,\ell]$ and denote the corresponding expectation by
$<\cdot>_{\ell}$. The two-point function in the infinite volume limit
$\ell \rightarrow \infty$ is then defined by
\begin{equation}
g(m-n) = <{\bf s}_m \cdot {\bf s}_n> = \lim_{\ell \rightarrow \infty}
<{\bf s}_m \cdot {\bf s}_n>_{\ell} \;\ge 0.
\end{equation}
If $m^*=0$, $g$ is independent of the boundary conditions.

To discuss the asymptotic decay of $g$ we first note that by
Griffiths second inequality $g$ is increasing in the couplings.
Thus
\begin{equation}
g(n) \ge {1 \over Z} \int \delta(|{\bf s}_0|-1)d^N s_0
\delta(|{\bf s}_n|-1)d^N {\bf s}_n \exp[\beta J(n){\bf s}_0 \cdot {\bf s}_n]
{\bf s}_0
\cdot {\bf s}_n
\end{equation}
which proves that $g(n)$ cannot decrease {\it faster} than the couplings
$J(n)$.

The {\it upper} bound for $g$ is slightly more complicated and uses the
well known Lieb-Simon type inequality. We define
\begin{equation}
\Lambda_L = \{u,v| \ {\rm either} \ |u| \le L, |v| > L \quad {\rm or}
\ |u| > L, \ |v| \le L \}
\end{equation}
and split the Hamiltonian as
\begin{eqnarray}
& \ & H_\lambda = H_1 + \lambda H_2, \nonumber \\
& \ & H_1 = - {1 \over 2} \hspace*{-  12 mm} \sum^{\ell}_{\hspace*{15 mm}
 m,n =
-\ell
\;\;\;  m,n \in \Lambda_L^c}  J(m-n) {\bf s}_m \cdot {\bf s}_n, \nonumber\\
& \ & H_2 = - {1 \over 2} \hspace*{-  12 mm} \sum^{\ell}_{ \hspace*{15 mm}
m,n = -\ell \;\;\; m,n \in \Lambda_L}
 J(m-n) {\bf s}_m \cdot {\bf s}_n.
\end{eqnarray}
Differentiating with respect to $\lambda$ we obtain
\begin{eqnarray}
& \ & <{\bf s}_m \cdot {\bf s}_n>_{\ell} = <{\bf s}_m \cdot {\bf s}_n>_{\ell,
\lambda=0}
+ \int^1_0 d \lambda {d \over {d \lambda}} <{\bf s}_m \cdot {\bf s}_n>_{\ell,
\lambda}
\nonumber \\ & \ & = \ <{\bf s}_m \cdot {\bf s}_n>_{\ell, \lambda=0} + 
\int^1_0 d
\lambda \; {\beta \over 2} \hspace*{-  12 mm}\sum^{\ell}_{\hspace*{12 mm} u,v =
-\ell \;\;\; u,v \in \Lambda_L} J(u-v)  \nonumber \\
& \ & (<({\bf s}_m \cdot {\bf s}_n)({\bf s}_u \cdot
{\bf s}_v)>_{\ell,\lambda}- <{\bf s}_m \cdot {\bf s}_n>_{\ell,\lambda}<{\bf s}_u \cdot
{\bf s}_v>_{\ell,\lambda}).
\end{eqnarray}
We choose $|m| \le L, |n| > L$. Then the first term in (2.6)
vanishes. For the second term we use the Gaussian domination valid
for $N=1,2,3,4$ \cite{B}
\begin{eqnarray}
<({\bf s}_m \cdot {\bf s}_n) ({\bf s}_u \cdot {\bf s}_v)>_{\ell, \lambda}
\le  <{\bf s}_m \cdot {\bf s}_n>_{\ell,\lambda}
<{\bf s}_u \cdot {\bf s}_v>_{\ell,\lambda} \nonumber \\
+ {1 \over N}(<{\bf s}_m \cdot {\bf s}_u>_{\ell,\lambda}<{\bf s}_m \cdot {\bf
s}_v>_{\ell,\lambda} + <{\bf s}_m \cdot {\bf s}_v>_{\ell,\lambda}<{\bf s}_n \cdot
{\bf s}_u>_{\ell,\lambda})
\end{eqnarray}
and set $\lambda =1$ because $<{\bf s}_m \cdot {\bf s}_n>_{\ell,\lambda}$
is increasing in $\lambda$. Finally we take $\ell \rightarrow \infty$
and arrive at
\begin{equation}
g(n) \le (\beta/N) \sum_{|u| \le L} \ \sum_{|v| > L} g(u) J(u-v)
g(v-n)
\end{equation}
for $|n| > L$.

The integral inequality (2.8) is studied in \cite{AN}. 
In essence, one splits the $v$-sum into terms with $|v| \le |n/2|$
and those with $|v| > |n/2|$. This yields for $|n|> \hat{L} \ , \ \hat{L}$
fixed,
\begin{equation}
g(n) \le c' |n|^{- \gamma} + \alpha(n) g(n/2)
\end{equation}
with $\alpha(n) \rightarrow 0$ as $|n| \rightarrow \infty$.
Iterating (2.9) results in a bound as  
\begin{equation}
g(n) \le c |n|^{- \gamma}.
\end{equation}
The precise conditions, cf.  \cite{AN}, Lemma 5.4, for the validity of (2.10) 
are (i) for $\gamma > 2$ it is required that $\lim_{n \to \infty} 
g(n) = 0$, which we know already from $m^{*}$ = 0, (ii) for $\gamma = 
2$ it is required that $g(n) \leq c_{1} (1 + \log(1+|n|))^{-1}$ with a 
suitable constant $c_{1}$ depending on the prefactor of $J$, (iii) for
$1 < \gamma <2$ it is required that $g(n) \leq c_{2}(1 + |n|)^{\gamma - 2}$
with a 
suitable constant $c_{2}$ depending on the prefactor of $J$.
We conclude that under the stated conditions on $g(n)$ and if $N=1,2,3,4$,
$m^* = 0$, then
\begin{equation}
g(n) \simeq \ {\rm const.} \ |n|^{-\gamma}
\end{equation}
for large $|n|$.

For Ising spins the bounds (2.3), (2.10) have recently been
sharpened \cite{NS}, such as to determine also the prefactor in (2.11).
Generalizing to the present case, we conjecture that
\begin{equation}
\lim_{n \rightarrow \infty} {1 \over {\beta J(n)}} <{\bf s}_0 \cdot 
{\bf s}_n> = {1 \over
\beta^2} N \chi^2.
\end{equation}
The proof in \cite{NS} uses the FK and percolation representation for the
lower bound and the random current representation for the upper
bound, which unfortunately are special to $N=1$.

\end{document}